\documentclass[sigconf,table]{acmart}

\usepackage{enumitem}
\usepackage[ruled,linesnumbered]{algorithm2e}
\usepackage{array,multirow}
\usepackage{amsmath}
\usepackage{algorithmwh}
\usepackage{footnote}
\makesavenoteenv{tabular}
\makesavenoteenv{table}
\usepackage{graphicx}

\copyrightyear{2020}
\acmYear{2020}
\setcopyright{acmlicensed}\acmConference[WebSci '20]{12th ACM Conference on Web Science}{July 6--10, 2020}{Southampton, United Kingdom}
\acmBooktitle{12th ACM Conference on Web Science (WebSci '20), July 6--10, 2020, Southampton, United Kingdom}
\acmPrice{15.00}
\acmDOI{10.1145/3394231.3397910}
\acmISBN{978-1-4503-7989-2/20/07}

\begin{document}

\title{Representativity Fairness in Clustering}

\author{Deepak P}
\email{deepaksp@acm.org}
\affiliation{%
  \institution{Queen's University Belfast, UK}
  \institution{Indian Institute of Technology Madras, India}
}
\author{Savitha Sam Abraham}
\email{savithas@cse.iitm.ac.in}
\affiliation{%
  \institution{Indian Institute of Technology Madras, India}
}

\begin{abstract}
Incorporating fairness constructs into machine learning algorithms is a topic of much societal importance and recent interest. Clustering, a fundamental task in unsupervised learning that manifests across a number of web data scenarios, has also been subject of attention within fair ML research. In this paper, we develop a novel notion of fairness in clustering, called representativity fairness. Representativity fairness is motivated by the need to alleviate disparity across objects' proximity to their assigned cluster representatives, to aid fairer decision making.  We illustrate the importance of representativity fairness in real-world decision making scenarios involving clustering and provide ways of quantifying objects' representativity and fairness over it. We develop a new clustering formulation, RFKM, that targets to optimize for representativity fairness along with clustering quality. Inspired by the $K$-Means framework, RFKM incorporates novel loss terms to formulate an objective function. The RFKM objective and optimization approach guides it towards clustering configurations that yield higher representativity fairness. Through an empirical evaluation over a variety of public datasets, we establish the effectiveness of our method. We illustrate that we are able to significantly improve representativity fairness at only marginal impact to clustering quality. 
\end{abstract}

\maketitle

\section{Introduction}\label{sec:intro}

Clustering is a classical unsupervised learning task~\cite{jain1988algorithms} that seeks to organize a dataset of objects in groups, such that objects that fall within the same group are more similar to each other than those that belong to different groups. Among the most popular formulations for clustering, inspired by the classical $K$-Means algorithm~\cite{macqueen1967some}, is that of centroid clustering. Such algorithms, in addition to grouping data objects into clusters, offer a {\it representative prototype} for each cluster. Within the classical $K$-Means formulation, the representative prototype for a cluster is simply the {\it centroid} of objects within the cluster. In other similar formulations such as $K$-medoids~\cite{rdusseeun1987clustering}, the representative could be the {\it medoid}, which is the object within a cluster whose average dissimilarity to all the objects in the cluster is minimal. $K$-Medoids may be more appropriate when the usage scenario is better suited towards looking at an {\it actual object} rather than a hypothetical object that is located at the centroid. In both these cases as well as several others, the target is to produce clusters along with a representative for each cluster that is {\it how centrally located} within the cluster. 

Consider a pragmatic way of using clustering within a real-world scenario as follows. For an oversubscribed job vacancy where it is infeasible to scan through each job application manually, clustering offers an easy way out. The employer would cluster these job applications into a moderate number of clusters, followed by looking at each cluster representative, make a decision on suitability (e.g., reject, shortlist or scrutinize further), and apply that decision to {\it all} objects within the respective cluster. In retail, a similar sequence, clustering of customers followed by manual appreciation of the cluster representatives, would aid identifying suitable customer clusters to offer a sales promotion to. Given that advanced data collection methods severely outpace any efforts on manually labelling each object, clustering benefits a plethora of modern scenarios involving large datasets. In fact, it is also very useful for simpler scenarios such as {\it facility location}. For example, a large retail company venturing into a new city could cluster potential customer data using their geo-locations and open branches at each cluster centroid to ensure proximity to potential customers. Across these myriad scenarios, the cluster representative would be consistently used to inform a decision that would be applied to all objects within the cluster. 

Looking back at the job screening scenario, consider a cluster that involves a large and diverse set of job applications. Large clusters are inevitable when the desired number of output clusters are much fewer than the total number of applications, as would often be the case (given the need to speed-up the job screening process). For a large cluster, some applications would inevitably be very close to the cluster representative, whereas other applications would be much further away. {\it A judgement or decision made through inspecting an object is intuitively likely to be more suited to objects that are very similar to it than to objects that are not as similar.} Given our setting where judgements and decisions are based on perusing the cluster representative, the judgement arrived at is likely to be better suited for applications very similar to the cluster representative, and are likely to be much less suitable for those that are much further away. Despite this disparity in suitability with varying similarities to the cluster representative, the same decision is applied to all job applications given the structure of the process. This opens up a frontier of potential unfairness in the process, which we call as {\it representativity (un)fairness}, since some applications are accorded judgements that are more suited than for others. It may be noted that this is directly influenced by the usage of the clustering technique than in the manual aspects of the process, and is thus best addressed within the clustering itself. 


\subsection{Clustering on the Web}

We have used job screening as a scenario to motivate representativity fairness primarily since this scenario has been the subject of much fairness oriented legislation\footnote{Examples include Title VII of the Civil Rights Act of 1964 (US), Uniform Guidelines on Employee Selection Procedures 1978 (US)}. However, the notion of fairness in general, as well as representativity fairness, are pertinent in any scenario involving exploratory data analysis using clustering methods. Web has emerged, over the past decades, as a rich source of (unlabelled) data. Thus, the web likely presents the largest set of scenarios involving exploratory analyses of data. Each user on the web leaves different cross-sections of digital footprints in different services she uses, together encompassing virtually every realm of activity. These data sources are so huge that no manual labelling efforts can keep pace with their growth. These service providers have an interest to perform exploratory analytics via clustering (clustering of mobility trajectories~\cite{yuan2017review}, for example) over consumer data for scenarios such as (i) understanding customer behavior to inform long-term corporate strategies, (ii) deliver personalized promotions and do better customer targeting for new services/products, and (iii) nudge individual users towards behavioral patterns that better suit their interests. In these scenarios, users who end up at the fringes of their assigned cluster, and thus are accorded low representativity, stand to be disadvantaged when decisions are made for them on the basis of their (distant and thus significantly dissimilar) cluster representative. These could induce a spectrum of scenarios, such as being offered irrelevant personalized offers or unsuitable sales promotions, to more consequential ones such as being offered an unfavorable car insurance quote. In the public sector, this could lead to more serious consequences. For example, users who end up on the perimeter of a cluster whose representative is considered typical of 'suspicious behavior' may be shortlisted for needless additional surveilance and/or pro-active checks. In short, it is easy to see how the massive amounts of data collected over the web infrastructure forms a fertile ground for clustering and exploratory analytics tasks, and why representativity could become a serious concern in scenarios within them. 



\subsection{Our Contribution} 

In this work, we develop a novel measure of fairness for the representative based clustering paradigm used across the $K$-Means inspired family of algorithms. In sharp contrast to a recent such work~\cite{chen2019proportionally} that considers unfairness at the level of object groups (they call those as {\it coalitions}), our new notion of fairness, that of {\it representativity fairness}, is based on (an aggregate of) individual object-level assessments. We demonstrate how representativity fairness can be facilitated within the clustering process itself through simple illustrative examples, and outline a number of evaluation measures to quantify representativity fairness of a clustering. We propose a clustering formulation and method to improve representativity fairness within clustering outputs. Through an empirical evaluation over multiple datasets, we illustrate the empirical effectiveness of our approach in generating clusters with significant improvements in representativity fairness, while only suffering marginal degradations in clustering quality over existing methods.

\section{Related Work}\label{sec:relwork}

We now briefly summarize a selection of some recent literature on; (1) fairness in machine learning, and (2) fair clustering algorithms. 

\subsection{Fairness in Machine Learning}

Fairness in machine learning has seen tremendous amounts of research interest over the past several years. The vast majority of fair ML work has focused on supervised learning, especially on classification~\cite{zafar2017fairness,huang2019stable}. Among tasks outside the traditional realm of supervised learning, fairness has been explored in retrieval~\cite{zehlike2017fa}, clustering~\cite{chierichetti2017fair} and recommendation scenarios~\cite{patro2019incremental}. 

Fairness in machine learning may be conceptualized using a number of different and mutually conflicting normative considerations~\cite{kleinberg2016inherent}. Two streams of notions, as introduced in~\cite{dwork2012fairness}, are those of {\it individual fairness} and {\it group fairness}. Individual fairness is focused on {\it consistent treatment} and strives to achieve configurations where {\it similar objects are assigned similar outcomes}. Group fairness, on the other hand, looks to ensure that outcomes be equitably distributed across data subgroups defined on {\it sensitive attributes} such as gender, race, ethnicity, nationality and religion. In other words, individual fairness may be regarded as focusing on the {\it process} whereas group fairness is evaluated on the {\it outcome}. As Sen alludes to in~\cite{sen2009idea}, these relate to the ideas of {\it niti} and {\it nyaya} respectively in classical Indian jurisprudence. Though individual and group fairness have been traditionally treated as distinct and conflicting within work on fair machine learning, this dichotomy has been subject to some recent criticism (refer~\cite{binns2020apparent}). 

\subsection{Fair Clustering}

Most work on fair clustering has focused on {\it group fairness}. Fair clustering algorithms endeavor to ensure some form of representational parity across groups defined on {\it sensitive attributes} in the output clusters. Sensitive attributes could include gender, ethnicity, nationality, religion or even age and relationship status when dealing with people datasets; broadly, any attribute on which fairness is sought to be achieved could be treated as a sensitive attribute. As a concrete example, when considering a single protected attribute, say {\it gender}, the group-fair clustering task is to ensure that {\it each cluster} has a gender ratio that is either identical to, or close enough to, the gender ratio across the whole dataset (or alternatively, a pre-specified ideal gender ratio). If such representational parity is achieved, downstream applications making use of the outputs of the clustering are likely to treat the groups defined on the protected attribute fairly. Techniques differ on whether group fairness ensuring mechanisms are modelled at the pre-processing stage (e.g.,~\cite{chierichetti2017fair}), optimization model (e.g.,~\cite{abraham2020fairness}) or as a post-processing step (e.g.,~\cite{bera2019fair}). Another facet of classifying fair clustering techniques is based on the flexibility to accommodate different numbers and types of sensitive attributes. While some techniques are designed to accommodate a single binary sensitive attribute~\cite{chierichetti2017fair,olfat2019convex}, some others can accommodate a single multi-valued attribute~\cite{ahmadian2019clustering,kleindessner2019fair}. There have also been techniques that can accommodate multiple sensitive attributes simultaneously~\cite{bera2019fair,abraham2020fairness}. A third facet of viewing fair clustering literature is whether the objective is to achieve theoretical fairness bounds~\cite{chierichetti2017fair}, or achieve empirical fairness improvements~\cite{ziko2019clustering,abraham2020fairness}. 

Fairness in clustering outside the framework of fairness over protected groups, such as the task we address in this paper, has been subject to limited exploration. In fact, to our best knowledge, there has been only one prior work in this direction. This recent work~\cite{chen2019proportionally} proposes a notion of {\it proportionality fairness} in clustering. Using the intuitive assumption that {\it individuals prefer to be closer to their cluster representative} (which we will also use in our formulation), the authors of~\cite{chen2019proportionally} define a new concept of {\it proportional clusterings}. Under their definition, a clustering solution may be regarded as proportional if there does not exist any set of at least $\left \lceil \frac{n}{k} \right \rceil$ data points ($n$ is the number of data objects in the dataset, and $k$ is the number of clusters) each of which would prefer the same particular data point to be their cluster representative, in preference to their currently assigned (separate) cluster representatives. This notion is extended to multiples of $\left \lceil \frac{n}{k} \right \rceil$ points as well. The proportionality notion disallows any group of $\left \lceil \frac{n}{k} \right \rceil$ proximal points to be split across multiple clusters even if it benefits the dataset-wide optimization; in a way, this is so since they are considered entitled to their own cluster. The authors illustrate that proportional clustering solutions may not always exist, propose a notion of approximate proportionality, and provide algorithms that can achieve theoretical guarantees of approximate proportionality in the output clusterings. While proportionality is built upon the same basic intuition of the assumed preference of data objects to be proximal to their cluster representative, it significantly differs from our notion of representativity fairness. Being closest to our task in spirit, we use this method as a baseline in our experiments. 

\section{Representativity Fairness}

We now outline the novel notion of fairness that we consider in this paper, that of {\it representativity fairness}. We discuss quantifying representativity and fairness over it, and outline representativity fairness enhancement by means of illustrative examples. 

\subsection{Quantifying Representativity}

Representativity of a data object within a specified clustering is the extent to which the data object is represented by the cluster representative corresponding to the cluster to which it is assigned. The clustering process makes use of a similarity measure between objects as a fundamental building block towards building clusters and cluster representatives. Thus, as a natural fallout, we also use similarity metrics to quantify representativity. Accordingly, the extent to which a data object is represented by it's cluster representative is simply the {\it similarity of the object to the cluster representative}. In other words, it is inversely related to the dissimilarity of the object to the cluster representative. The dissimilarity of an object to it's assigned cluster representative may be seen as the cost incurred by the object due to the cluster-level abstraction provided by the clustering. 

\subsection{Quantifying Representativity Fairness}\label{sec:quantrepfair}

Our notion of representativity fairness is rooted on the concept of egalitarianism, and seeks to achieve egalitarianism on representativity. Thus, {\it we would prefer clusterings where objects fare equally well on representativity}. In other words, an ideal configuration for representativity fairness would be the case where all objects are equidistant from their respective cluster representatives. This enforces that all objects should live on the surface of equal-sized hyperspheres centered on their respective cluster representatives. This is evidently an infeasible scenario for many datasets since there may not exist $k$ cluster representativies where all data objects live on the surface of the equal-radius hyperspheres centered on them. Thus, we need to be able to quantify clusterings based on the extent to which they adhere to the notion of representativity fairness. Consider a dataset $\mathcal{X} = \{ \ldots, X, \ldots \}$ and a clustering $\mathcal{C} = \{ \ldots, C, \ldots\}$ where $C$ represents a cluster. Let $R(C)$ represent the representative of cluster $C$, and $\mathcal{C}(X)$ denote the cluster to which $X$ belongs under the clustering $\mathcal{C}$. Thus, the representativity of objects in $\mathcal{X}$ under the clustering $\mathcal{C}$ is given by the set/distribution:

\begin{equation}\label{eq:repvector}
    \mathcal{R}(\mathcal{X}, \mathcal{C}) = \{\ dist(X, R(\mathcal{C}(X))) \ |\ X \in \mathcal{X}\ \}
\end{equation}

Our intent, given our target of egalitarianism, is to ensure that the values within $\mathcal{R}(\mathcal{X}, \mathcal{C})$ are {\it as even as possible}. A natural first way to quantify this is by means of the {\it variance} of the distribution:

\begin{equation}\label{eq:var}
    Var(\mathcal{R}(\mathcal{X}, \mathcal{C})) = \frac{1}{|\mathcal{X}|} \sum_{X \in \mathcal{X}} \bigg(\mathcal{R}[X] - avg\{\mathcal{R}[X])|X \in \mathcal{X}\}\bigg)^2
\end{equation}

where $\mathcal{R}[X]$ is a shorthand for $dist(X, R(\mathcal{C}(X)))$. The more {\it representativity fair} a clustering is, the lower the value of $Var(\mathcal{R}(\mathcal{X}, \mathcal{C}))$. Resource allocation in distributed systems has a similar structure as representativity 'allocation' in clustering, and a fairness notion that was developed for the latter~\cite{jain1984quantitative} is intuitively appealing and appropriate for our setting. The measure, often referred to as the {\it Jain} measure from the name of the first author, offers a score in the range $(0,1]$ with higher values indicating higher fairness:

\begin{equation}\label{eq:jain}
Jain(\mathcal{R}(\mathcal{X}, \mathcal{C})) = \frac{\bigg( \sum_{X \in \mathcal{X}} \mathcal{R}[X] \bigg)^2}{|\mathcal{X}| \times \sum_{X \in \mathcal{X}} (\mathcal{R}[X])^2}
\end{equation}

For a perfectly uniform distribution (say, $\{ 2, 2, 2 \}$, across three objects), the numerator and denominator both evaluate to the same value (in this case, $36$), yielding a $Jain$ of $1.0$. Any deviations from perfect uniformity with the same sum/budget do not matter to the numerator (since it is a function of the sum), but increase the denominator value, thus causing $Jain$ to drop from $1.0$ to lower values, approaching $0.0$ for highly asymmetric distributions over large $|\mathcal{X}|$ settings. It may be noted that variance or $Jain$ do not capture the absolute values of $\mathcal{R}[X]$s, but simply the uniformity. Thus, there could be cases where a low variance is achieved within a configuration where the cluster representative is very far away from all cluster members. In view of preventing such undesirable cases, we would additionally want to consider the average of the $\mathcal{R}(\mathcal{X}, \mathcal{C})$ as an evaluation measure. Note that $Avg$ (which is essentially the normalized sum) is the objective that many clustering algorithms directly or indirectly try to optimize for. Turning our attention back to variance, quantifying representativity fairness using variance incentivizes moving towards what is often understood as {\it strict egalitarianism}\footnote{https://plato.stanford.edu/entries/justice-distributive/\#Strict} on representativity which penalizes deviations on both directions from the mean equally. Thus, a clustering that penalizes a small minority of points' representativity for higher representativity for a large majority could still fare reasonably well on variance, $Jain$ and average. Theories of justice have, over the decades, developed notions that prefer some deviations from strict egalitarianism over others. One example is a philosophy called {\it luck egalitarianism}~\cite{arneson2004luck} which argues that inequalities be justified as long as they benefit people who are victims of bad luck. A simpler and high-level philosophy put forward in a classical work by Rawls~\cite{john1971theory} that has come to be known as the {\it difference principle} suggests that {\it inequalities be arranged to the greatest benefit of the least advantaged}. Reflections of this Rawlsian position are also found in Gandhian thought and the Indian constitution~\cite{sharma1989rawlsian}. Inspired indirectly by these, we consider the representativity of the object accorded least representativity (i.e., highest distance from cluster representative) as another complementary measure to evaluate representativity fairness:


\begin{equation}\label{eq:max}
    Max(\mathcal{R}(\mathcal{X}, \mathcal{C})) = max\{\ \mathcal{R}[X] \ |\ X \in \mathcal{X}\ \}
\end{equation}

The lower the values of each of $Var$ and $Max$ (while keeping $Avg$ low as well) and higher the value of $Jain$, the more representativity fair the clustering would be. As noted earlier, a typical fairness-agnostic clustering algorithm such as classical $K$-Means would be expected to naturally optimize for $Avg$; thus, a fairness-conscious algorithm (such as the one we develop in this paper) would be expected to trade-off $Avg$ while seeking to achieve lower values on $Var$ and $Max$ and correspondingly higher values on $Jain$. 

\begin{figure}
  \includegraphics[width=1.8in]{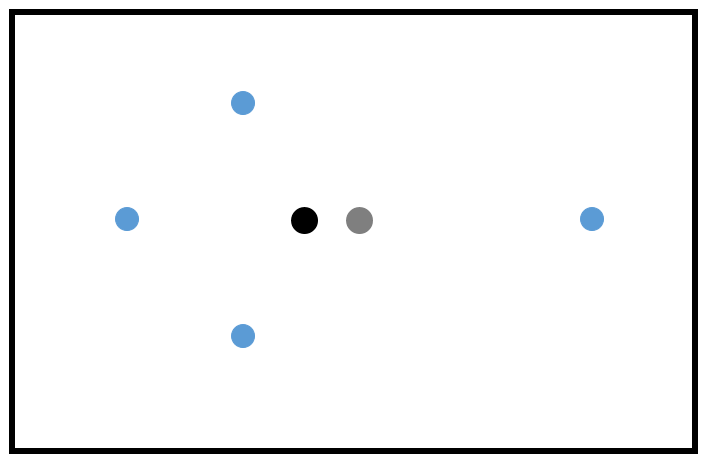}
  \caption{Single Cluster Example (best viewed in color)}
  \label{fig:toy1}
\end{figure}

\begin{table}[tbp]
\begin{tabular}{|c|c|c|}
\hline
{\bf Object} & {\bf Distance to} & {\bf Distance to} \\
& {\bf Black (2.5,2)} & {\bf Grey (3,2)} \\
\hline
(2,1) & {\bf 1.12} & 1.41 \\
(1,2) & {\bf 1.50} & 2.00 \\
(2,3) & {\bf 1.12} & 1.41 \\
(5,2) & 2.50 & {\bf 2.00} \\
\hline
$Avg$ & {\bf 1.56} & 1.71 \\
\hline
$Max$ & 2.50 & {\bf 2.00} \\
\hline
$Var$ & 0.43 & {\bf 0.11} \\
\hline
$Jain$ & 0.88 & {\bf 0.97} \\
\hline
\end{tabular}
\caption{Analysis of Figure~\ref{fig:toy1}}
\label{tab:toy1}
\end{table}

\subsection{Representativity Fairness: Examples}

We now motivate, by means of examples, as to how representativity fairness could be facilitated by varying (i) {\it construction of cluster representatives}, or (ii) {\it cluster memberships of data objects}. The method that we develop in a later section targets to enhance representativity fairness by affecting both kinds of variations. 

\subsubsection{Varying Cluster Representatives}\label{sec:varyreps}

Consider a single cluster comprising the {\it blue} objects/points in Fig~\ref{fig:toy1}, with the data point co-ordinates themselves listed in Table~\ref{tab:toy1}. The centroid of the blue points, which also corresponds to the best estimate to minimize $Avg$ for the cluster, is represented by the black point (at $(2.5,2)$ as outlined in Table~\ref{tab:toy1}). The $\mathcal{R}[X]$ values when considering the black point as the cluster representative is indicated in the second column along with corresponding $Avg$, $Max$, $Var$ and $Jain$ values. While being the centroid of all blue points, it may be noted that the black point offers much lower representativity to the blue point on the far right, given the configuration of the other blue points. Now, consider an alternative cluster representative for the same set of blue points, indicated by the grey point (at $(3,2)$). The $\mathcal{R}[X]$ values as well as $Avg$, $Max$, $Var$ and $Jain$ are indicated in the third column in the table. It is easy to note that changing the cluster representative from the black point to the grey point improves the representativity of the blue point in the far right, by enhancing the proximity of the cluster representative to itself. This is reflected in the analysis in Table~\ref{tab:toy1} that offers a comparative perspective between the two settings for cluster representatives, with the better value in each row indicated in boldface. The choice of the grey point as a cluster representative is seen to offer better values of $Max$, $Var$ and $Jain$ with improvements of $20\%$, $74\%$ and $10\%$ respectively as compared to the choice of the black point, while trailing the latter on the $Avg$ measure by $10\%$. The grey point also enables achieving a very high $Jain$ value, very close to the upper bound of $1.0$. This illustrates that representativity fairness may be enhanced by deviating from the $K$-Means paradigm of a centrally located cluster representative. 

\begin{figure}
  \includegraphics[width=0.9\columnwidth]{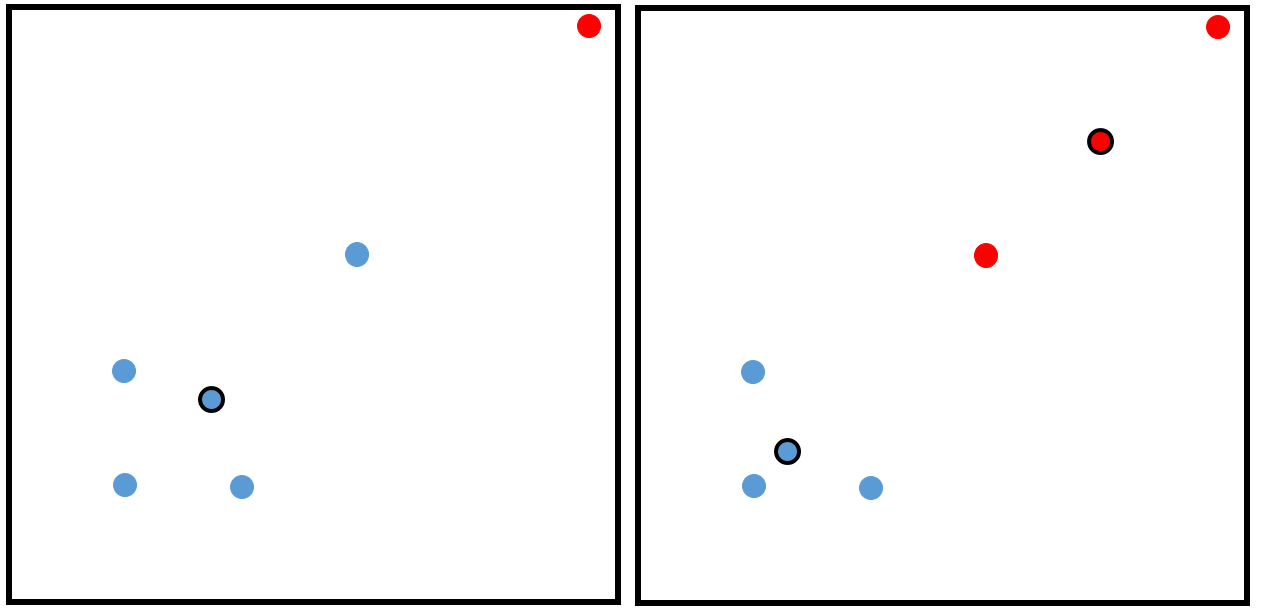}
  \caption{Two Clusters Example (best viewed in color)}
  \label{fig:toy2}
\end{figure}

\begin{table}[tbp]
\begin{tabular}{|c|c|c|}
\hline
 & {\bf Left} & {\bf Right} \\
\hline
$\mathcal{R}[(1,1)]$ & 1.06 & {\bf 0.47} \\
$\mathcal{R}[(1,2)]$ & 0.79 & {\bf 0.74} \\
$\mathcal{R}[(2,1)]$ & 0.79 & {\bf 0.74} \\
$\mathcal{R}[(3,3)]$ & 1.77 & {\bf 1.41} \\
$\mathcal{R}[(5,5)]$ & {\bf 0.00} & 1.41 \\
\hline
$Avg$ & {\bf 0.88} & 0.96 \\
\hline
$Max$ & 1.76 & {\bf 1.41} \\
\hline
$Var$ & 0.40 & {\bf 0.19} \\
\hline
$Jain$ & 0.71 & {\bf 0.86} \\
\hline
\end{tabular}
\caption{Analysis of Figure~\ref{fig:toy2}}
\label{tab:toy2}
\end{table}

\subsubsection{Varying Cluster Memberships}\label{sec:varycluster}

We now use another example to illustrate how representativity fairness can be facilitated by varying cluster memberships. Towards illustrating that this is independent of Section~\ref{sec:varyreps}, we will stick with the $K$-Means paradigm of {\it centroid as cluster representative} for this example. Figure~\ref{fig:toy2} shows a dataset clustered in two different ways, with cluster membership indicated using color coding; all points, blue and red, are data points, with cluster centroids represented using a black ring. In the clustering on the left, all but one data point is part of the blue cluster, and the lone element on the top right is in a red cluster (since the red cluster centroid would overlap with that data point, it is not shown to avoid confusion). The figure on the right has a different configuration for the two clusters, one where the top right point is joined in the red cluster by another point (with the red cluster's centroid at the midpoint between the two points). An analysis, similar to earlier, follows on Table~\ref{fig:toy2}. While the left configuration, a potential stable $K$-Means output, scores better than the right configuration on the $Avg$ measure by around $9\%$, the right configuration comfortably beats the left on the $Max$ (by $20\%$), $Var$ (by $53\%$) and $Jain$ ($21\%$) measures, and may rightly be regarded as being more adherent to representativity fairness. In fact, the right configuration also offers better representativity for $4$ out of $5$ points in the dataset. This example illustrates that representativity fairness can be facilitated by changes in cluster memberships, in addition to changing the cluster representative as seen earlier. 

\subsection{Positioning Representativity Fairness}\label{sec:positioning}

We now analyze representativity fairness within the context of the two streams of fairness, viz., individual and group fairness, as outlined in Section~\ref{sec:relwork}. Our definition of representativity fairness, due to not using the notion of protected groups, may be considered as being unrelated to group fairness. However, it does deviate from the general notion of individual fairness notably. As an example, consider the object $(3,3)$ in Section~\ref{sec:varycluster}. It's proximity to each of $\{(1,1),(1,2),(2,1)\}$ is equal to or better than it's proximity to $(5,5)$. Yet, assigning it to the same cluster as $(5,5)$, in sharp dsiagreement to the {\it `similar objects be assigned similar outcomes (i.e., cluster memberships)'} tenet of individual fairness, yields better representativity fairness to the overall clustering. In a way, representativity fairness incentivizes outcomes that are better for those objects that are disadvantaged in terms of their neighborhood density. We believe that neighborhood density disadvantage would also be correlated with minority/non-mainstream groups, and thus, we expect representativity fairness to be imbibing a flavor of group fairness in practice despite not using groups within the definition. The connection between proportionality~\cite{chen2019proportionally} and representativity fairness is more nuanced. Proportionality is focused towards being fair to those objects who collectively prefer a new cluster representative; thus, violations of proportionality occur more on the fringes of clusters much like where objects disadvantaged on representativity also appear. However, there are sharp contrasting elements between the two notions due to being motivated by different scenarios. Within the framework of representativity fairness, individual cluster members can cause deterioration in representativity fairness based on how far they are positioned from the cluster representative. On the other hand, no proportionality violations are regarded as taken place until a strong enough collective voice (note that the cardinality of the set is an important consideration in proportionality) emerges from across the fringes of multiple neighboring clusters; thus, proportionality does not necessarily prevent an individual object from having very low representativity. Thus, representativity fairness accounts for quasi-outliers whereas proportionality does not bear that flavour. 

\section{Problem Definition}

As outlined in Section~\ref{sec:quantrepfair}, let $\mathcal{X} = \{ \ldots, X, \ldots\}$ be a dataset of objects which are defined over a set of attributes $\mathcal{A} = \{ \ldots, A, \ldots\}$ that are deemed to be pertinent for the clustering task. Much like in the $K$-Means family of methods, we model the distance between any two objects defined over $\mathcal{A}$ as:

\begin{equation}
    d(X, Y) = \sum_{A \in \mathcal{A}} \big( X[A] - Y[A] \big)^2
\end{equation}

where $X[A]$ is the value that object $X$ takes on attribute $A$. $d(.,.)$ is the square of the euclidean distance, and thus, is monotonically related to the euclidean distance, which we denote as $dist(.,.)$ as in Section~\ref{sec:quantrepfair}. The task of clustering is that of partitioning $\mathcal{X}$ into a pre-specified number of clusters or groups, denoted as $\mathcal{C} = \{ \ldots, C, \ldots\}$; as in convention in literature, we use $k$ to denote the pre-specified number of clusters to be formed in the output. The clustering $\mathcal{C}$ is expected to be comprehensive over $\mathcal{X}$, with each object $X$ being assigned a unique cluster, denoted by $\mathcal{C}(X)$. Additionally, we expect each cluster to be associated with a cluster representative, modeled as an object over the same space of attributes $\mathcal{A}$. As outlined in Section~\ref{sec:quantrepfair}, we use $R(C)$ to denote the representative for the cluster $C \in \mathcal{C}$. For ease of reference, we use the term clustering to refer to both the grouping provided by $\mathcal{C}$ as well as the set of cluster representatives associated with the clusters in $\mathcal{C}$. 

The conventional task of clustering targets to achieve a grouping that maximizes intra-cluster similarity and minimizes inter-cluster similarity, similarity being inversely related to the distance as defined above. The task of {\it representativity fair clustering}, on the other hand, intends to obtain a clustering that yields better values on $Var$ (Eq.~\ref{eq:var}), $Jain$ (Eq.~\ref{eq:jain}) and $Max$ (Eq.~\ref{eq:max}) measures as evaluated on the representativity vector $\mathcal{R}(\mathcal{X}, \mathcal{C})$ (Eq.~\ref{eq:repvector}). While gains on this is expected to be achieved at the cost of deterioration in the $Avg$, keeping such deterioration small would be considered better. 

\section{RFKM: Our Method}

We now describe our proposed method for representativity fairness in clustering, which we call {\it RFKM} to stand for both {\it R}representativity {\it F}airness and {\it K}-{\it M}eans, the latter being the method that it draws inspiration from. {\it RFKM} incorporates two novel loss terms that collectively incentivize reducing variability among representativity of objects within the output clustering as well as towards ensuring attention towards objects that are accorded the least representativity. The {\it RFKM} objective function is as follows:

\begin{multline}
    \mathcal{O} = \sum_{X \in \mathcal{X}} d(X, R(\mathcal{C}(X))) + \lambda_1 \times \sum_{X \in \mathcal{X}} \big(d(X, R(\mathcal{C}(X)))\big)^2 \\
    \hspace{1in} + \lambda_2 \times max\{d(X, R(\mathcal{C}(X))) | X \in \mathcal{X}\}
\end{multline}

This objective function has three components. The first term is the usual $K$-Means objective that computes the loss based on the sum of each data object's dissimilarity/distance to its cluster representative; given that higher distances offer lower representativity, we refer to these loss terms as object-level representativity loss. The second term is modelled as the sum of squares of the object-specific representativity losses. The squaring operation amplifies the larger losses more than smaller ones, and thus optimizing for the squared versions would help nudge the clustering towards those that minimize the larger losses. The third term is simply the representativity loss corresponding to the object that is farthest from its cluster representative. The overall objective is modelled as a weighted sum, with $\lambda_1$ and $\lambda_2$ being the weights for the second and third terms that may be set appropriately. As in $K$-Means, the overall loss is computed over a {\it given} clustering; the RFKM task is thus to identify a clustering that minimizes this overall loss. 

\subsection{Intuitive Basis of the Loss Terms}\label{sec:intuitive}

The first loss term, being the classical $K$-Means loss, is more aligned with lowering $Avg$ and targets to lower the sum of the representativity losses. The second term deepens the attention within the optimization formulation towards those objects that have high representativity losses. This may be seen as offering a push towards lower $Var$ {\it from above downward}; there is no corresponding term to push from below since we would ideally like to achieve low $Var$ at low overall representativity losses as well, and the constraints from the geometry of the data offer a natural upward push. These would entail better distributional parity targeted at yielding higher $Jain$ scores. The third term measures the representativity loss associated with the object that is placed farthest from its cluster representative, in the clustering. This is directly targeted towards optimizing for $Max$. However, the second term also helps reducing $Max$ (and albeit less so, the first term too) since the farthest object also forms the largest term within the first and second summations. 

\subsection{The Optimization Approach}

Clustering is a dataset-wide optimization problem, and the $K$-Means formulation yields an NP-hard problem~\cite{vattani2009hardness}. It follows that the RFKM objective is also NP-hard. Thus, much like the case of $K$-Means, we adopt an iterative optimization approach that, while susceptible to local minima, optimizes for the objective gradually across iterations. Notice that there are two sets of variables; (i) the cluster memberships, and (i) the cluster representatives. We adopt the high-level $K$-Means framework of optimizing for each of these in turn (which, as one may notice, correspond to the $E$ and $M$ steps of the classical expectation-maximization meta-algorithm~\cite{dempster1977maximum}). 

One difficulty with the RFKM objective is the construction of the third term; we replace it with a differentiable approximation:

\begin{multline}\label{eq:maxapprox}
    max\{d(X, R(\mathcal{C}(X))) | X \in \mathcal{X}\} \\ \approx \frac{1}{\phi} log_e \bigg( \sum_{X \in \mathcal{X}} exp(\phi \times d(X, R(\mathcal{C}(X))))\bigg)
\end{multline}

where $\phi$ is a sufficiently large positive number (say, 10). This approximation is easy to understand if one notices that the multiplication with $\phi$ and the exponentiation enlarge the largest representativity loss much more than the others (for larger $\phi$, the larger the amplification). Thus, the sum of such enlarged losses are dominated by the largest loss, which is then recovered through the $log(.)$ and division by $\phi$. For smaller values of $\phi$ (say, close to $1$), this approximation would exceed the {\it max}, as it may be intuitive to see; this construction allows for some control to amplify the importance of the largest representativity loss further. Thus, the overall objective may be re-written as:

\begin{multline}
    \mathcal{O} = \sum_{X \in \mathcal{X}} d(X, R(\mathcal{C}(X))) + \lambda_1 \times \sum_{X \in \mathcal{X}} \big(d(X, R(\mathcal{C}(X)))\big)^2 \\
    \hspace{1in} + \frac{\lambda_2}{\phi} log_e \bigg( \sum_{X \in \mathcal{X}} exp(\phi \times d(X, R(\mathcal{C}(X))))\bigg)
\end{multline}

While the summation over $\mathcal{X}$ offers convenient interpretation, the optimization is better understood if it is written equivalently as a summation over clusters, as follows:

\begin{multline}\label{eq:rfkmobj}
    \mathcal{O} = \sum_{C \in \mathcal{C}} \sum_{X \in C} d(X, R(C)) + \lambda_1 \times \sum_{C \in \mathcal{C}} \sum_{X \in C} \big(d(X, R(C))\big)^2 \\
    \hspace{1in} + \frac{\lambda_2}{\phi} log_e \bigg( \sum_{C \in \mathcal{C}} \sum_{X \in C} exp(\phi \times d(X, R(C)))\bigg)
\end{multline}

\subsubsection{Cluster Representative Updates.} While just optimizing for the first term (as in $K$-Means) allows achieving a closed-form solution for estimating a new cluster representative given the cluster memberships, the inclusion of the other loss terms cause much dependencies. However, the first derivative can be equated to zero to give an iterative update formula as follows:

\begin{multline}\label{eq:clusterrepupdate}
    \forall A \in \mathcal{A}, R(C)[A] = \\
    \frac{\sum\limits_{X \in C} X[A] \bigg(1 + 2 \lambda_1 d(X,R(C)) + \frac{\lambda_2 \times exp(\phi\ d(X,R(C)))}{\sum\limits_{C' \in \mathcal{C}}\sum\limits_{X' \in C'} exp(\phi\ d(X',R(C')))} \bigg)}{\sum\limits_{X \in C} \bigg(1 + 2 \lambda_1 d(X,R(C)) + \frac{\lambda_2 \times exp(\phi \ d(X,R(C)))}{\sum\limits_{C' \in \mathcal{C}}\sum\limits_{X' \in C'} exp(\phi \ d(X',R(C')))} \bigg)}
\end{multline}

where $R(C)[A]$ denotes the value associated with the $A^{th}$ attribute of the cluster representative. It is easy to note that this equation is {\it not} in closed form, given that what is to be estimated, i.e., $R(C)$, also appears on the RHS of the equation. It may further be seen that all cluster representatives, $R(C'), \forall C' \in \mathcal{C}$ affect each other (note the denominator of the third term in the numerator as well as denominator). The intuitive appeal for Eq.~\ref{eq:clusterrepupdate} is evident when it is viewed as estimating each cluster representative as a weighted sum of objects in the cluster with object-specific weights, with object-specific weights being directly related to:

\begin{itemize}[leftmargin=*]
    \item the current representativity loss of the object; \&
    \item an amplified and normalized version of the current representativity loss of the object across all attributes. 
\end{itemize}

This construction drags the cluster representative closer to objects that are currently much farther away from itself. Relating this to the example in Figure~\ref{fig:toy1}, this would move the cluster representative from the position of the black object further towards the grey object due to the far right object being accorded much more weight due to it being far away as compared to the others. It may also help to note that for sufficiently large values of $\phi$, the third term approximates as follows:

\begin{multline}
    \frac{\lambda_2 \times exp(\phi \times d(X,R(C)))}{\sum_{C' \in \mathcal{C}}\sum_{X' \in C'} exp(\phi \times d(X',R(C')))} \\
    \approx 
    \begin{cases}
    \lambda_2 & \text{if $X$ is the object that is most distant from} \\ & \text{its currently assigned cluster representative} \\
    0 & \text{otherwise}
    \end{cases}
\end{multline}

Thus, for every object other than the one with the highest representativity loss, the third term becomes negligible. 

\subsubsection{Cluster Assignment Updates.} Each of the first two terms in the objective have one term corresponding to each object, with the third term relating to the entire clustering. Under the current estimates of cluster representatives, the cluster assignment can be varied to set it to what would yield the smallest value for the overall objective; this yields:

\begin{multline}\label{eq:clusterassign}
    \mathcal{C}(X) = \mathop{\arg\min}_{C \in \mathcal{C}} \bigg( d(X,R(C)) + \\ 
    \lambda_1\ (d(X,R(C)))^2 + \lambda_2 \mathcal{T}_{Edit(\mathcal{X},\mathcal{C},X,C)} \bigg)
\end{multline}

where $\mathcal{T}_{Edit(\mathcal{X},\mathcal{C},X,C)}$ is the term $\frac{1}{\phi} log_e \big( \sum_{C' \in \mathcal{C}} \sum_{X' \in C'} exp(\phi \times d(X, R(C')))\big)$ evaluated over the clustering $\mathcal{C}$ of the dataset $\mathcal{X}$, with a single change in cluster assignment, that of re-assigning $X$ to $C$; we do not change the cluster representative during the course of this re-assignment. Recall from Eq.~\ref{eq:maxapprox} that $\frac{1}{\phi} log_e \big( \sum_{C' \in \mathcal{C}} \sum_{X' \in C'} exp(\phi \times d(X, R(C')))\big)$ is an approximation for the maximum representativity loss according to clustering $\mathcal{C}$ over $\mathcal{X}$. Thus, the third term in Eq.~\ref{eq:clusterassign}, in effect, adds an incentive to favour cluster assignments that reduce the max representativity loss. In other words, the third term introduces significant resistance to cluster assignment updates that would increase the max representativity loss across the clustering and vice versa.




While this does not follow that RFKM can cause a direct change from the left configuration to the right configuration in Figure~\ref{fig:toy2}, the RFKM objective scores $27\%$ worse for the left configuration than the right (in contrast, the $K$-Means objective prefers the left configuration); this entails that RFKM would prefer to gravitate towards the right configuration as compared to the left one, across iterations. 

\begin{algorithm}
\caption{\bf \textit{RFKM}}
\begin{flushleft}
Input. Dataset $\mathcal{X}$, Attribute Set $\mathcal{A}$, number of clusters $k$ \\
Hyper-parameters: $\lambda_1$ and $\lambda_2$, {\it max iterations} \\
Output. Clustering $\mathcal{C}$ and associated cluster representatives
\end{flushleft}
\begin{code}
1. {\it Initialize $k$ clusters} \\
2. {\it Set cluster representatives using Eq.~\ref{eq:clusterrepupdate}} \\
3. $while(not\ yet\ converged\ and\ max.\ iterations\ not\ reached)$ \\
4. \> $\forall X \in \mathcal{X}$, \\
5. \> \> {\it Set C(X) using Eq.~\ref{eq:clusterassign}} \\
6. \> \> {\it Update cluster representatives as outlined in Eq.~\ref{eq:clusterrepupdate}} \\
8. {\it Return }$\mathcal{C}${\it  along with the cluster representatives}
\end{code}
\label{alg:rfkm}
\end{algorithm}

\subsection{The Overall Technique}

Having detailed the separate steps of the optimization process, we are now ready to summarize the overall approach. As outlined in Algorithm~\ref{alg:rfkm}, RFKM starts with a random initialization of cluster memberships followed by iterative refinement alternating between re-estimating cluster memberships and cluster representatives. 

\subsubsection{Complexity.} Let the number of objects, attributes, clusters and the maximum number of iterations be $n$, $m$, $k$ and $t$ respectively. Then, the complexity of RFKM is in $\mathcal{O}(nmkt)$, making it asymptotically as fast as $K$-Means. The key point to note is that within the cluster representative learning step in Eq.~\ref{eq:clusterrepupdate}, the denominator in the third term of the object weight construction is independent of the choice of the current cluster, i.e., $C$. Thus, it can be pre-computed before each cluster assignment separately, and used to weigh the contribution from each object within the cluster. Given the linear cost, RFKM compares favorably against recent fair clustering algorithms that are super-quadratic~\cite{chierichetti2017fair} and quadratic~\cite{abraham2020fairness}.

\begin{table}[tbp]
\begin{tabular}{|c|c|c|c|}
\hline
{\bf Name} & {\bf \# Instances} & {\bf \# Attributes} & {\bf \# Classes} \\
\hline
Iris & 150 & 4 & 3 \\
\hline
Yeast & 1484 & 8 & 10 \\
\hline
Wireless\footnote{short for Wireless Indoor Localization} & 2000 & 7 & 4 \\
\hline
Avila & 20867 & 10 & 12 \\
\hline
Letter\footnote{short for Letter Recognition} & 20000 & 16 & 26 \\
\hline
\end{tabular}
\caption{Dataset Statistics}
\label{tab:dataset}
\end{table}

\section{Experimental Evaluation}

We now describe the empirical evaluation of our method against existing clustering formulations. We start by describing the datasets and experimental setup, followed by baselines and evaluation measures. We present results of empirical analyses across a number of real-world datasets, and analyze the results across various facets. 

\subsection{Datasets and Experimental Setup}\label{sec:datasets}

We use a number of datasets from the UCI Machine Learning repository~\cite{Dua:2019} for our empirical study. The usage of public datasets, we hope, will aid benchmarking and reproducibility. The details of the datasets we employ in our study are listed in Table~\ref{tab:dataset}. These incorporate a wide variety of dataset sizes, ranging from $150$ to $20k$, a wide range of attribute numbers ($4$ to $16$) and a range of number of classes ($3$ to $26$). Given that clustering is an unsupervised learning task, the class labels are only used for evaluation. In all cases, unless otherwise mentioned, we set the number of desired output clusters, i.e. $k$, to the number of classes. This is intuitive since we are interested in analyzing whether clustering is capable of capturing the class-wise grouping of objects. The choice of datasets with varying sizes and attributes are intended to illustrate the generalizability of our empirical study. We set the value of $\lambda_1$ to $1.0$; we will study the trends against varying values of $\lambda_1$ separately. The third term in Eq.~\ref{eq:rfkmobj} is an approximation of the {\it max}, and thus, this term would be expected to be quite small when compared with the first two that sum over all objects; accordingly, we set $\lambda_2$, its co-efficient to be $\frac{|\mathcal{X}|}{10}$ to ensure it is well-accounted. We empirically observed that values for $\phi = 3$ is sufficient to achieve a good approximation of $max$, and thus, set it thus. 

\subsection{Baselines and Evaluation Measures}

\subsubsection{Baselines.} Our main baseline is the recent fair clustering work~\cite{chen2019proportionally} that also builds its fairness measure upon proximity and representation, albeit in a significantly different way. There are two techniques that they propose, which we will refer to as $LS$ (for {\it Local Search}) and $Greedy$ respectively, within our experimental analyses. Much like the empirical evaluation in~\cite{chen2019proportionally}, we also compare our approach to the $K$-means method (denoted as $KM$). We also use the same heuristic for cluster initialization in line 1 of RFKM (refer Algorithm~\ref{alg:rfkm}). Given the random initialization step in $KM$ and $RFKM$, we consistently report the average results over $100$ random initializations, for robustness. 

\subsubsection{Evaluation Measures.} Much like the structure used in~\cite{abraham2020fairness}, we would like to evaluate the RFKM clusterings on two fronts; (i) representativity fairness, and (ii) clustering quality. As developed in Section~\ref{sec:quantrepfair}, we will use $Var$, $Jain$ and $Max$ for measuring representativity fairness. For clustering quality, we use the following measures:

\begin{itemize}[leftmargin=*]
    \item {\it Avg (for K-Means Objective):} The $K$-Means objective measures the coherence of clusters by way of aggregating the distances of each object to its cluster representative. It may be noted that $Avg$, the measure discussed in Section~\ref{sec:quantrepfair}, is related to the per capita $K$-Means objective, i.e., $K$-Means objective normalized by the dataset size. It is also notable that $K$-Means objective is the only clustering quality evaluation measure used in~\cite{chen2019proportionally}. 
    \item {\it Silhouette Score (Sil)}: Silhouette~\cite{rousseeuw1987silhouettes} measures the separatedness of clusters, and quantifies a clustering with a score in $[-1,+1]$, higher values indicating well-separated clusters. This was used in~\cite{abraham2020fairness} as a clustering quality metric. 
    \item {\it Clustering Purity (Pur):} Yet another way to measure the quality of the clustering is to see how well it adheres to the manual labellings available in the dataset. Clustering purity\footnote{https://nlp.stanford.edu/IR-book/html/htmledition/evaluation-of-clustering-1.html} is a popular measure that captures the alignment between clusters and dataset labels:
    \begin{equation}
    Pur(\mathcal{C},\mathcal{L}, \mathcal{X}) = \frac{1}{|\mathcal{X}|}\sum_{C \in \mathcal{C}} \mathop{\max}_{L \in \mathcal{L}} |C \cap L|
    \end{equation}
    It may however be noted that some of the datasets that we use are designed for classification benchmarking; Thus, we do not expect clustering methods to deliver very high purities over them. Still, the relative trends across the methods would offer a legitimate comparative perspective. 
\end{itemize}
It may be noted that higher values are desirable on $Sil$, $Pur$ and $Jain$, whereas lower values are desirable on all other measures. 


\begin{table*}[tbp]
\begin{tabular}{|c|r|r|r|r|r|r|r|r|r|r|r|r|}
\hline
\hline
{\it Dataset} & \multicolumn{4}{c|}{{\it Var} $\downarrow$} & \multicolumn{4}{c|}{{\it Jain} $\uparrow$} & \multicolumn{4}{c|}{{\it Max} $\downarrow$} \\ \cline{2-13}
{\it Name} & {\it LS} & {\it Greedy} &  {\it KM} & {\it RFKM} & {\it LS} & {\it Greedy} &  {\it KM} & {\it RFKM} & {\it LS} & {\it Greedy} &  {\it KM} & {\it RFKM} \\ 
\hline
\hline
{\it Iris} & 0.23 & 0.27 & {\bf 0.11} & {\bf 0.11} & 0.75 & 0.69 & 0.80 & {\bf 0.81} & 2.42 & 2.56 & 1.66 & {\bf 1.63} \\
{\it Yeast} & 1.21E-2 & 1.63E-2 & 6.66E-3 & {\bf 6.26E-3} & 0.71 & 0.72 & 0.84 & {\bf 0.86} & 0.86 & 0.87 & 0.77 & {\bf 0.70} \\
{\it Wireless} & 33.27 & 61.43 & 19.96 & {\bf 17.83} & 0.84 & 0.77 & 0.84 & {\bf 0.88} & 44.61 & 53.41 & 35.84 & {\bf 31.20} \\
{\it Avila} & 10.95 & \cellcolor{gray!10} & 1.62 & {\bf 1.07} & 0.25 & \cellcolor{gray!10} & 0.64 & {\bf 0.78} & 403.64 & \cellcolor{gray!10} & 43.00 & {\bf 14.28} \\
{\it Letter} & 3.73 & \cellcolor{gray!10} & 2.88 & {\bf 2.33} & 0.91 & \cellcolor{gray!10} & 0.93 & {\bf 0.95} & {\bf 16.74} & \cellcolor{gray!10} & 17.81 & {\bf 16.74} \\
\hline
\hline
RFKM Perf. & \multicolumn{4}{c|}{{\bf 18.64\% better}} & \multicolumn{4}{c|}{{\bf 5.68\% better}} & \multicolumn{4}{c|}{{\bf 46.81\% better}} \\ 
\hline
\end{tabular}
\caption{Representativity Fairness Evaluation. Notes: (i) Arrows next to measures indicate whether higher or lower values are desirable. (ii) The best value for each measure on each dataset is highlighted in bold. (iii) Some runs of the Greedy approach did not complete in reasonable amounts of time and memory, and thus, those cells are greyed out. }
\label{tab:repfaireval}
\end{table*}

\begin{table*}[tbp]
\begin{tabular}{|c|r|r|r|r|r|r|r|r|r|r|r|r|}
\hline
\hline
{\it Dataset} & \multicolumn{4}{c|}{{\it Avg} $\downarrow$} & \multicolumn{4}{c|}{{\it Sil} $\uparrow$} & \multicolumn{4}{c|}{{\it Pur} $\uparrow$} \\ \cline{2-13}
{\it Name} & {\it LS} & {\it Greedy} &  {\it KM} & {\it RFKM} & {\it LS} & {\it Greedy} &  {\it KM} & {\it RFKM} & {\it LS} & {\it Greedy} &  {\it KM} & {\it RFKM} \\ 
\hline
\hline
{\it Iris} & 0.83 & 0.78 & {\bf 0.65} & 0.68 & 0.49 & 0.51 & {\bf 0.55} & {\bf 0.55} & 0.90 & {\bf 0.95} & 0.89 & 0.89 \\
{\it Yeast} & {\bf 0.18} & 0.20 & 0.19 & 0.19 & 0.10 & 3.68E-3 & 0.26 & {\bf 0.27} & 0.32 & 0.32 & {\bf 0.42} & 0.41 \\
{\it Wireless} & 13.22 & 14.18 & {\bf 10.46} & 11.13 & 0.33 & 0.25 & {\bf 0.40} & 0.39 & 0.88 & 0.82 & {\bf 0.93} & 0.90 \\
{\it Avila} & 1.89 & \cellcolor{gray!10} & {\bf 1.70} & 1.92 & 0.03 & \cellcolor{gray!10} & 0.15 & {\bf 0.18} & 0.41 & \cellcolor{gray!10} & {\bf 0.46} & 0.42 \\
{\it Letter} & {\bf 6.07} & \cellcolor{gray!10} & 6.41 & 6.47 & 0.08 & \cellcolor{gray!10} & {\bf 0.15} & {\bf 0.15} & 0.05 & \cellcolor{gray!10} & {\bf 0.16} & {\bf 0.16} \\
\hline
\hline
RFKM Perf. & \multicolumn{4}{c|}{{\bf 7.20\% behind}} & \multicolumn{4}{c|}{{\bf 4.27\% better}} & \multicolumn{4}{c|}{{\bf 4.12\% behind}} \\ 
\hline
\hline
\end{tabular}
\caption{Clustering Quality Evaluation. Notes: (i) Arrows next to measures indicate whether higher or lower values are desirable. (ii) The best value for each measure on each dataset is highlighted in bold. (iii) Some runs of the Greedy approach did not complete in reasonable amounts of time and memory, and thus, those cells are greyed out. }
\label{tab:clusqualeval}
\end{table*}


\subsection{Experimental Results}

We now analyze the comparative performance of RFKM against {\it KM}, {\it LS}~\cite{chen2019proportionally} and {\it Greedy}~\cite{chen2019proportionally} on the two fronts; representativity fairness and clustering quality. 

\subsubsection{Representativity Fairness.} The representativity fairness evaluation appears on Table~\ref{tab:repfaireval}. As expected, RFKM consistently performs better than the competing techniques on each of $Var$, $Jain$ and $Max$. The per-measure aggregate improvements, the average of row-specific percentage improvements, are recorded at the bottom row. Between $Var$ and $Max$, the performance improvements are much higher for $Max$ as against $Var$ and $Jain$. While RFKM is targeted to optimize for all three measures, it is easier to rein in the few high values in the $\mathcal{R}[.]$ vector than to reduce dispersion across all; this reflects in the high improvements recorded for $Max$. Among $Var$ and $Jain$, the latter has an upper bound of $1.0$, and with some baselines, values being already beyond $0.80$, there is {\it `not enough space'} to improve, unlike the case of $Var$. In fact, RFKM records a $4.6$ percentage point improvement on the $Jain$ measure which is significant and substantial in those ranges. The improvements, while consistent, differ across datasets. The quantum of improvements are quite small for the $Iris$ dataset; this is likely because $Iris$ has just $150$ data points spread across $4$ attributes and $3$ classes. This provides limited possibilities in arriving at alternative clusterings that optimize for representativity fairness while still retaining cluster coherence. That $KM$ fares ahead of $LS$ and $Greedy$ may be considered as an empirical indication that the notion of proportionality that $LS$ and $Greedy$ use is reasonably different from the notion of representative fairness that we evaluate. 

\subsubsection{Clustering Quality Evaluation.}\label{sec:clusterqualeval} 
As indicated in Section~\ref{sec:quantrepfair}, we expect that clusterings that seek to advance representativity fairness are likely to take a hit on clustering quality metrics, given that there these criteria are not necessarily at harmony with each other. The clustering quality as evaluated over $Avg$, $Sil$ and $Pur$ are outlined in Table~\ref{tab:clusqualeval}. True to expectations, RFKM records a better performance on these metrics only on a minority of scenarios; in particular, RFKM is the top performer on only $5$ combinations\footnote{RFKM is joint best on 2 out of those combinations.} out of $15$ ($5$ datasets, $3$ clustering quality measures). We will first analyze the performance on $Avg$ and $Pur$. RFKM is seen to lag $7.2\%$ and $4.12\%$ behind the next best performing method on the $Avg$ and $Pur$ measures respectively. It may however be noted that the next best performing method is not always the {\it 'same'} method; in certain cases, it is $KM$ and it is $LS$ and $Greedy$ in certain other cases. That said, given that $KM$ is an overwhelming frontrunner (scoring highest in $10$ out of $15$ combinations), a straight comparison pitting RFKM against $KM$ would evaluate to a $4.97\%$ deterioration on $Avg$ and $3.14\%$ deterioration on $Pur$. These deteriorations are seen to be quite limited, and quite small when compared to the gains achieved on representativity fairness. Turning our attention to $Sil$, RFKM records a different picture. RFKM is seen to be performing better than the baselines quite consistently on $Sil$, and records an average of $4.27\%$ improvement. While this indeed be regarded as surprising, the cluster representative learning step in RFKM offers some cues to explain this result. $Sil$ measures how well separated the cluster representatives are, with respect to the objects in the dataset. The cluster representative learning step in RFKM accords higher weighting to {\it far off} data objects, dragging the representative towards them. To ensure meaningful movement, over iterations, it is plausible that different cluster representatives be dragged in different directions, enhancing their mutual separation. Such effects are likely behind the better RFKM performance on $Sil$, and these observations point to interesting future work as to the use of representativity fairness in more general scenarios that focus on particular aspects of clustering quality that are aligned with $Sil$. 

\begin{figure}
  \includegraphics[width=\columnwidth]{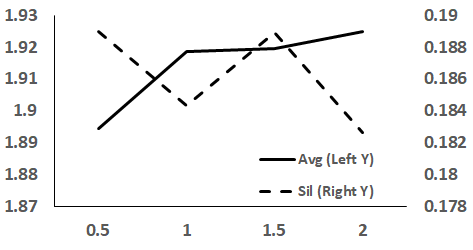}
  \caption{Avila: $Avg$ and $Sil$ vs. $\lambda_1$}
  \label{fig:avgsil}
\end{figure}

\begin{figure}
  \includegraphics[width=\columnwidth]{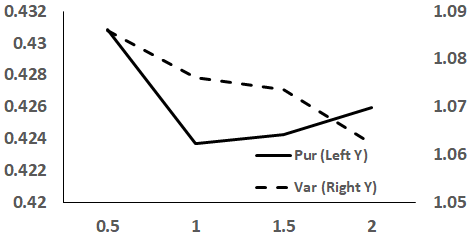}
  \caption{Avila: $Pur$ and $Var$ vs. $\lambda_1$}
  \label{fig:purvar}
\end{figure}

\begin{figure}
  \includegraphics[width=\columnwidth]{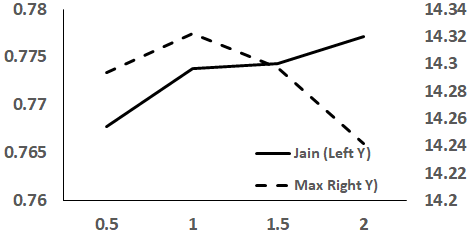}
  \caption{Avila: $Jain$ and $Max$ vs. $\lambda_1$}
  \label{fig:jainmax}
\end{figure}

\subsubsection{Trends with $\lambda_1$} 

We now analyze the RFKM performance against the various measures, varying the value of $\lambda_1$ from $0.5$ to $2.0$ in steps of $0.5$. $\lambda_1$ is a way of setting the strength of the second term in Eq.~\ref{eq:rfkmobj}, the term that strives to reduce the larger representativity losses towards ensuring lower $Var$ and higher $Jain$ scores. The second term is also related to $Max$, though that is more directly handled by the third term. Thus, with increasing $\lambda_1$, we would expect {\it higher values of Jain and Avg}, and {\it lower values on Var, Max, Sil and Pur}. In other words, we would expect {\it better values on representativity fairness measures} and {\it poorer values on clustering quality measures}. We observed consistent trends across the datasets, and plot only the {\it Avila} dataset trends here for brevity. With each of these measures being in different ranges, we plot these across Figures~\ref{fig:avgsil},~\ref{fig:purvar} and~\ref{fig:jainmax}, each figure plotting two measures, one on the {\it left Y} axis and another on the {\it right Y} axis; the legend indicates whether the measure is plotted on the left or right. Across these plots, we observe consistent, gradual and expected trends on $Avg$ (increase recorded in Fig~\ref{fig:avgsil}), $Var$ (decrease recorded in Fig~\ref{fig:purvar}) and $Jain$ (increase recorded in Fig~\ref{fig:jainmax}). $Sil$ and $Pur$ are seen to be swinging within very small ranges (Fig~\ref{fig:avgsil} and Fig~\ref{fig:purvar}) and may be regarded as stable within these ranges of $\lambda_1$. On the other hand, $Max$ shows consistent and expected decrease except for the case of $\lambda_1=0.5$, as seen in Fig~\ref{fig:jainmax}. Overall, these illustrate that the second term broadly works in line with the intuition in Sec~\ref{sec:intuitive}. It is also interesting to note that the evaluation measures are fairly smooth on $\lambda_1$ and do not cause any abrupt changes. 

\section{Conclusions}

We considered the task of fairness in clustering and oultined motivating scenarios where disparities in similarities with cluster representatives could lead to decisions that vary in the degree of appropriateness across data objects. Based on this, we developed a novel notion of fairness, called {\it representativity fairness}, and outlined measures of quantifying it. We sketched ways of enhancing representativity fairness by way of examples, and developed a novel clustering formulation, RFKM, that builds upon classical $K$-Means to optimize for representativity fairness. RFKM incorporates novel loss terms and uses an alternating iterative optimization approach to optimize for the combination of loss terms in the objective. Through an extensive empirical evaluation over a variety of real-world datasets over appropriate baselines, we illustrate that RFKM is able to achieve significant gains on representativity fairness at very limited impact on clustering quality. 

\subsection{Future Work}

We are considering two different directions of future work in representativity fairness. First, as outlined in Section~\ref{sec:clusterqualeval}, we are exploring ways to tease out the relationship between representativity fairness and silhoutte scores, towards developing newer insights that could inform fair clustering research. Second, we are looking into common clustering formulations and their treatment of minority groups in the dataset, with an eye on seeing whether they are correlated with lower representativity fairness, as discussed in Section~\ref{sec:positioning}. 

\bibliographystyle{ACM-Reference-Format}
\bibliography{repfairness}

\end{document}